# GENERATION AND PARAMETERIZATION OF FORCED ISOTROPIC TURBULENT FLOW USING AUTOENCODERS AND GENERATIVE ADVERSARIAL NETWORKS

Kanishk[1], Tanishk Nandal[1], Prince Tyagi[1], Raj Kumar Singh[1]

[1]Delhi Technological University, New Delhi, India

## ABSTRACT

*Autoencoders and generative neural network models have recently gained popularity in fluid mechanics due to their spontaneity and low processing time instead of high fidelity CFD simulations. Auto encoders are used as model order reduction tools in applications of fluid mechanics by compressing input high-dimensional data using an encoder to map the input space into a lower-dimensional latent space. Whereas, generative models such as Variational Auto-encoders (VAEs) and Generative Adversarial Networks (GANs) are proving to be effective in generating solutions to chaotic models with high 'randomness' such as turbulent flows. In this study, forced isotropic turbulence flow is generated by parameterizing into some basic statistical characteristics. The models trained on pre-simulated data from dependencies on these characteristics and the flow generation is then affected by varying these parameters. The latent vectors pushed along the generator models like the decoders and generators contain independent entries which can be used to create different outputs with similar properties. The use of neural network-based architecture removes the need for dependency on the classical mesh-based Navier-Stoke equation estimation which is prominent in many CFD softwares.*

## 1. INTRODUCTION

The advancements in computational and simulation capabilities in recent years have led to better than ever analysis of complex fluid flows. This in turn has led to vast amounts of multi-dimensional data representing the intricate details of these chaotic flows. Due to the challenges posed by the vastness of this ever-increasing data in analysing complex numerical simulation problems, extraction of important physical parameters from this data is indispensable. Characteristics of such high-fidelity DNS data can be contained in a low dimensional latent subspace. Model reduction is used to extract coherent key features and reduce data and computation costs. Several Model Order Reduction techniques have been employed and refined to analyse flows with high spatial and temporal complexity.

Various traditional model reduction techniques including proper orthogonal reduction, dynamic mode decomposition, global stability analysis, Koopman analysis, and Resolvent analysis have been utilized to analyse fluid flows [1]. POD and DMD are based on Singular value decomposition. POD is generally used along with projection techniques such as Galerkin Projection. However, it is found that Galerkin projection models are unstable under many circumstances [2,3], and this sparks interest to look into non-projection-based approaches. Also, the rapid advancements in machine learning and especially neural networks have accelerated the study of fluid flows and have shown great potential in modelling and study of these flows due to their handling of nonlinear phenomena. Hence it is not surprising that various reduced-order modelling techniques based on machine learning (ML-ROMs) have recently been proposed [5-7]. Neural network-based dimensionality reduction techniques often employ autoencoders.

In recent years, there have been many developments in the field of deep learning due to the availability of massive computing power. Variational autoencoders [8] are generative models that encode the input as a probabilistic distribution. Since their proposal, they have been consistently leveraged. For instance, Gregor et al. [9] combined an attention mechanism with a sequential variational auto-encoding framework that generates highly realistic images indistinguishable from the naked eye. Other enhanced models such as Importance Weighted Autoencoders [10] and Ladder Autoencoders [11] have also been presented. Another variant of autoencoders that has been utilized heavily for engineering problems is convolutional autoencoders. Convolutional autoencoders differ from traditional autoencoders due to weight sharing. Hence, they have been demonstrated to work well and are extensively used in image processing

 

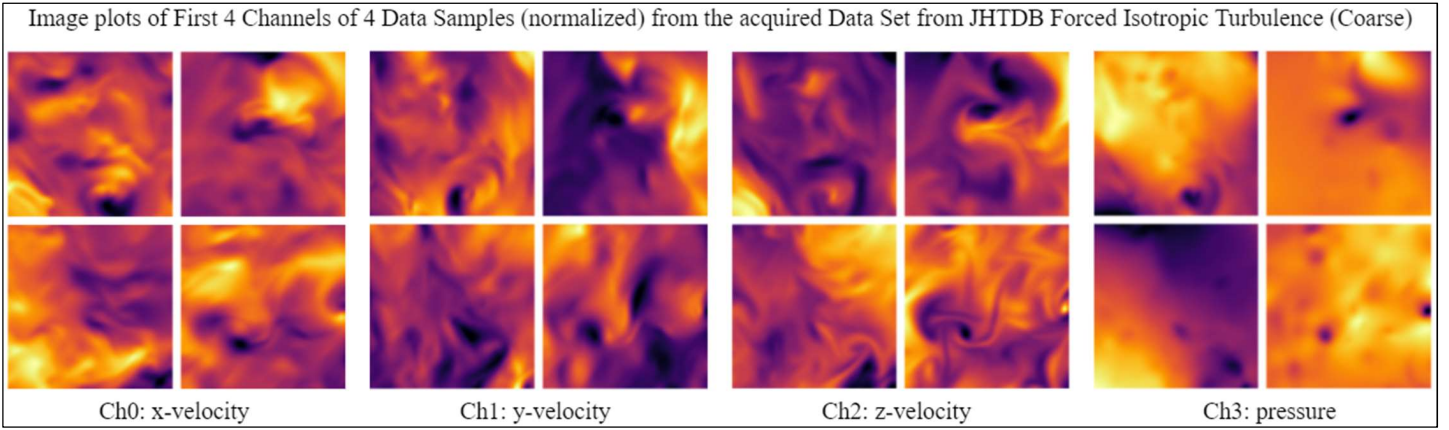

**FIGURE 1** EXTRACTED DATA SAMPLES FROM THE ORIGINAL DATASETS

applications. Maski et al. [12] proposed the idea of stacked CAEs for hierarchical feature extraction. Du et al. [13] proposed a model by stacking denoising autoencoders in a convolutional manner. Research for advancements in autoencoders is thus ongoing.

Autoencoders have proved to be of great use in the order reduction of fluid flows. Fukami et al. [14] presented the assessment of an autoencoder which consisted of a convolutional neural network and multi-layer perceptron and evaluated the model based on parameters like number of latent modes, choice of activation function, and number of weights. Mohan et al. [15] proposed a physics-embedded convolution-based autoencoder (PhyCAE), by enforcing the incompressibility of the fluid flow within CNNs. Srinivasan et al. [16] demonstrated the potential of LSTM to predict turbulent statistics. Similarly, Nakamura et al. [17] presented a CNN-based AE combined with LSTM to predict the temporal turbulent flow. Cheng et al. [18] proposed a hybrid deep adversarial autoencoder by combining a variational autoencoder with a GAN for nonlinear flow prediction. With the proposal of these different autoencoder models, there arises a need to compare how accurately and efficiently do these different autoencoders encode and predict various fluid data. This study aims to parameterize a forced isotropic turbulence flow into statistical characteristics of turbulence by testing different models of convolutional and variational autoencoders along with Generative Adversarial Network. This study also aims to create models for reconstruction and generation of the turbulent flow from the extracted parameters to steer the turbulence modelling techniques towards neural network based systems instead of rigorous mathematical formulations and derivations.

## 2. MATERIALS AND METHODS

### 2.1 Dataset and Data preprocessing

The dataset used in training the models is from the JHTDB Forced Isotropic Turbulence Simulation [25,26,27]. 2400 64x64 random slices from the total of 1024x1024x1024 cells are extracted containing flow values in various channels. The extracted data is of the size 2400x16x64x64. The 16 channels contain values of x, y, z velocities (u, v, w), pressure (p), gradients ($u_x$, $u_y$, $u_z$, $v_x$, $v_y$, $v_z$, $w_x$, $w_y$, $w_z$) and pressure gradients ($p_x$, $p_y$, $p_z$). These are selected in order to extract their dependencies on each other and help in estimation of the flow parameters and eventually help in reconstruction and generation of the flow after passing it through the proposed Neural Network models.

The simulation parameters as given in [27] are as following:
Simulation Domain: $2\pi$ x $2\pi$ x $2\pi$
Viscosity ($\nu$) = 0.000185
Simulation time-step $\Delta t$ = 0.0002
$\delta t$ = 0.002
Time step stored between t=0 and 10.056: (5028 time samples separated by $\delta t$)

There are 9 characteristic flow parameters that are calculated for individual data slices of 16x64x64. Following characteristic parameters were calculated for each slice ($\nu$ = 0.000185 $E_{k_{int}}$=0.40864) [27]:

1. Total Kinetic Energy: $E_{tot} = \frac{1}{2}\langle u_i u_i \rangle$
2. Dissipation Factor: $\varepsilon = 2\nu\langle S_{ij}S_{ij}\rangle$, where $S_{ij}$ are the gradient values of velocity
3. RMS Velocity: $u' = \sqrt{\frac{2}{3}E_{tot}}$
4. Taylor Micro Scale: $\lambda = \sqrt{15\,\nu\,u'^2/\varepsilon}$
5. Taylor-Scale Reynolds No.: $R_\lambda = u'\lambda/\nu$
6. Kolmogorov time scale: $\tau_n = \sqrt{\frac{\nu}{\varepsilon}}$
7. Kolmogorov length scale: $\eta = \nu^{3/4}\varepsilon^{-1/4}$
8. Integral scale: $L = \frac{\pi}{2u'^2}E_{k_{int}}$
9. Large eddy turnover time: $T = L/u'$

These parameters are stored for each data slice and forms a vector of 2400x9. This matrix is normalized using min-max normalization and the values are mapped between -1 and 1 for the integration with the models. Same goes for the main dataset as well.

## 2.2 Autoencoders

Autoencoders are neural networks that are trained to match their output to the input. An autoencoder that passes the input through a code with dimensions less than the input is called undercomplete autoencoder (Figure 1). It has two parts: An encoder function $z = f(x)$ that encodes the input to a latent space and a decoder function $r = g(z)$ that reconstructs the input from this latent space and maps it to the output. The bottleneck layer forces the autoencoder to learn the most useful properties of the input. It is trained simply by minimizing the loss function $\mathcal{L}(x, g(z))$ such as MSE:

$$\mathcal{L}_{MSE}(x, g(z)) = \frac{1}{n} ||g(z) - x||_2^2 \qquad (1)$$

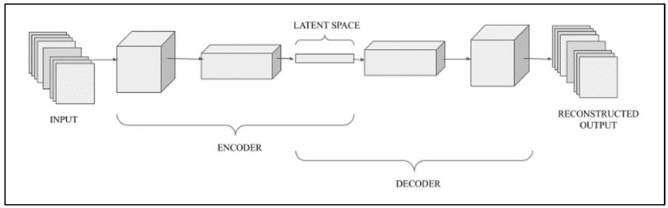

**FIGURE 2**: BASIC REPRESENTATION OF AN UNDERCOMPLETE AUTOENCODER

### 2.2.1 Variational Autoencoders

A variational autoencoder is a stochastic generative model that consists of a probabilistic decoder given by a likelihood function $p_\theta(x|z)$ and an encoder $q_\varphi(z|x)$ which approximates the posterior $p_\theta(z|x) \propto p_\theta(x|z)p(z)$ which is intractable, where $p(z)$ is a prior distribution over latent $z$. The decoder and encoder learn simultaneously by maximizing the variational lower bound, i.e.,

$$\mathcal{L}(x; \theta, \varphi) = \mathbb{E}_{q(Z|X)} \log p(x|z) - D_{KL}(q(z|x)||p(z)) \qquad (2)$$

This equation represents the combination of reconstruction loss and the KL divergence of the approximate posterior and the prior of the latent variable $z$.

### 2.2.2 Convolutional Autoencoders

Convolutional Autoencoders combine the traditional convolutional networks with an autoencoder architecture. The encoder consists first of convolutional and max pooling layers as described earlier. To further reduce dimensions, the tensor is then reshaped and fed into a series of linear layers that lead up to the latent layer.

The decoder then maps this latent layer to the output via a series of deconvolutional (transposed convolutional) or upsampling layers that upsamples the code to the original dimensions. In our model, upsampling layers were used as the deconvolutional layers seemed to introduce artifacts. The reconstructed output is checked against a loss function to minimized the per pixel reconstruction loss.

## 2.3 Generative Models

To generate new types of flows with the same initialization parameters, Generative models are used. Generative models are generally used to create new data instances with the help of some initialization variables and/or some predetermined inputs. They tend to capture the probability of the transformation of the inputs to the desired output and generate new data instances based on the probability distributions in the input vector space.

Here, we are aiming to generate new flow distributions for a certain set of flow parameters extracted from the JHTB dataset. We are aiming to utilize Conditional Deep Convolutional Generative Adversarial Networks with and without relativistic discrimination and Pix2Pix networks (as described here, [1]) with and without relativistic discrimination.

### 2.3.1 Conditional GAN

Generative Adversarial Networks usually have two models, the Generator (G) and the Discriminator (D). Both the Generator and the Discriminator are nonlinear functions like a multi-layer perceptron or a CNN, which maps the inputs to respective outputs.

In a classic GAN, G usually receives inputs **z** and tries to map the probability of getting the data **x**, D takes the mapped output of G and classifies it as to whether the output comes from the data **x** or generated by G. The aim is to train a G such that D makes a lot of mistakes in identifying the generated data from the original. [22]

The generator and the discriminator models are trained together and the aim is to adjust the parameters of G such that the (1- loss of D with the generated output of G) is to be minimized and the loss of D with real data points is also minimized at the same point.

In Conditional Adversarial Networks, an extra information **y** is fed to both G and D. Both are conditioned on the extra information **y**. Let the generator is fed a noise of the distribution $p_z(z)$, along with **y**. While training, G tries to maximize the following expression, and D tries to minimize it:

$$\mathbb{E}_x[\log(D(x|y))] + \mathbb{E}_z[\log(1 - D(G(z|y)))] \qquad (3)$$

In the model that is adopted for the generation of flows, the input **y** is a vector containing a normalized array of flow parameters that were extracted by processing individual instances of elements from the main data set **x**.

The generator G is made of both hidden layers as well as a CNN-based upscaling model for image creation. The G gets **y** and **z** as the input, where **z** is the latent variable from a normal distribution with a mean of 0 and a standard deviation of 1. The discriminator D gets **y** and either the image slice **x** or G(z|y) which is the generator output as its inputs and outputs a single Float value as the validity indicator.

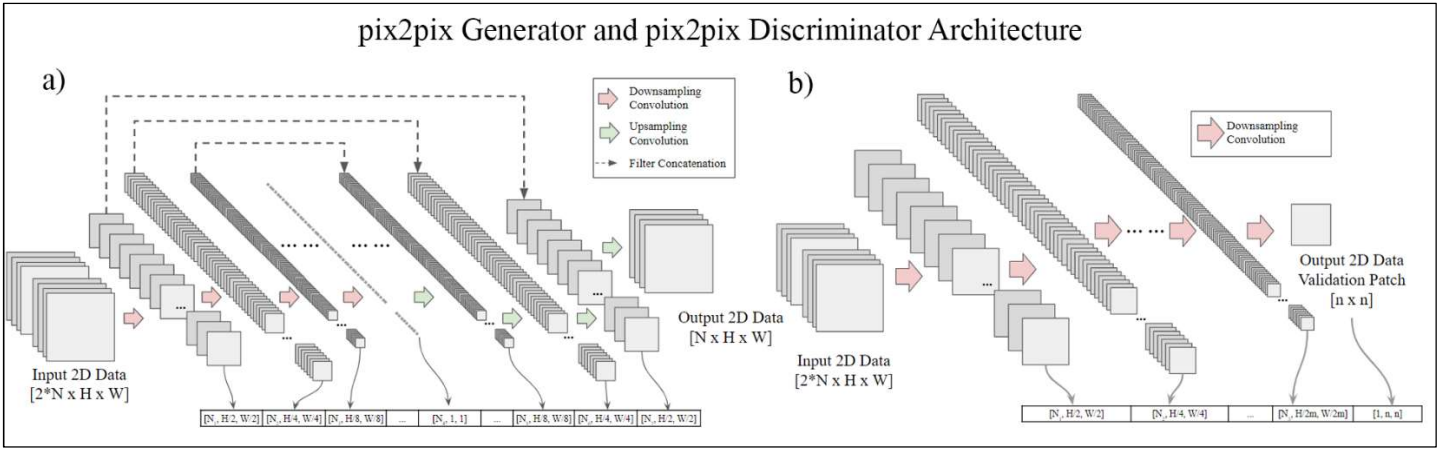

**FIGURE 3**: PIX2PIX ARCHITECTURE. A) DESCRIBES GENERATOR ARCHITECTURE ALONG WITH THE LAYERS THAT ARE ARRANGED TO FORM THE OUTPUT. B) DESCRIBES THE DISCRIMINATOR. N, W, H REPRESENTS THE NUMBER OF CHANNELS, THE WIDTH AND THE HEIGHT IN THE INPUT IMAGE DATA RESPECTIVELY.

In our implementation, y is replaced by parameters p which along with some normally distributed noise z are used as input for the generator G to produce 'fake' data. The Generator loss can be expressed as:

$$L_G = \mathbb{E}_{adv}[D(G(z|p),p),V] \quad (4)$$

where V is a validity vector of ones as its elements and $\mathbb{E}_{adv}$ represents the adversarial error function which is a mean Squared Error function in our case. $L_G$ is then backpropagated through G using the Adam optimizer at a suitable learning rate.

As for Discriminator D, the fake and real losses can be expressed as:

$$V_{real} = D(x,p) \quad (5)$$
$$V_{fake} = D[G(z|p),p] \quad (6)$$
$$L_{real} = \mathbb{E}_{adv}(V_{real},V) \quad (7)$$
$$L_{fake} = \mathbb{E}_{fake}(V_{fake},F) \quad (8)$$
$$L_D = \frac{1}{2}(L_{real} + L_{fake}) \quad (9)$$

where $F$ is a zero vector to account for 'truly fake' data. $L_D$ is then backpropagated through D using the Adam optimizer to modify the model parameters. Both G and D run against eachother and should settle to a stable $L_G$ and $L_D$ values to appropriately train the model.

### 2.3.2 Pix2Pix Architecture

Convolution-based image to image generative networks have become immensely popular nowadays. They usually consist of a Convolutional network that is trained to 'morph' the input 2D image similar to the desired output. pix2pix is one such GAN architecture [23]. Pix2pix is generally used for style transfer, image colorization, label identification, image reconstruction from edge maps, etc. Here we are using this approach to translate a randomly generated noise map to generate turbulent flows.

The pix2pix implementation in the previously mentioned reference is based upon 'U-Net' architecture [24]. The U-Net approach utilizes skip connections in between convolutional layers to retain the previously encoded layers.

Let the input image instance be x, let z be the random noisy input and y be the desired output image instance that the generator G needs to map x and z towards y or G: {x, z} → y. The Discriminator is also trained parallelly to distinguish the real images from the fake or the generated ones.[24] The objective function of such conditional GAN can be described as

$$\mathcal{L}_{cGAN}(G,D) = \mathbb{E}_{x,y}[\log(D(x|y))] \\ + \mathbb{E}_{x,z}[\log(1 - D(x|\,G(x|z)\,))] \quad (10)$$

where G is trying to minimize the above objective function and D is trying to maximize it. An unconditioned variant in which the discriminator doesn't observe x is also taken into consideration as

$$\mathcal{L}_{GAN}(G,D) = \mathbb{E}_y[\log(D(y))] + \mathbb{E}_{x,z}[\log(1 - D(G(x|z)))] \quad (11)$$

From experimentation, it was concluded that if we use L2 loss, the data output was not very clear to comprehend and most of its details were lost in doing so. A direct L1 loss relation was utilized to produce more clear data to 'fool' the discriminator:

$$\mathcal{L}_{L1}(G) = \mathbb{E}_{x,y,z}\left[||y - G(x|z)||_1\right] \quad (12)$$

The final objective function becomes,

$$G^* = \arg\min_G \max_D \mathcal{L}_{cGAN}(G,D) + \lambda \mathcal{L}_{L1}(G) \quad (13)$$

In our implementation, we are using a similar model structure as the one given in original reference [23], which has a CNN based generator with skip layers in between.

For each instance of training, two sets of data are created, one with the original data set **x** and other **x'** which is **x + z_n** where $z_n$ is a 2D noise distribution. Generator also receives some external parameters which like data can be written as **p** and **p'** as the unmodified and modified parameters respectively. The generator output can be written as G(x|p) or G(x'|p') depending upon the inputs given. The discriminator D is straightforward another neural network-based model which receives two data instances and itself gives out an n x n dimensional array.

Let the modified data instance be **x'**, **p'** and the unmodified data be **x**, **p** and let these variables represent the 'real' data. Then the overall GAN loss and pixelwise loss can be written as

$$L_{GAN} = \mathbb{E}_{GAN}[D(G(x'|p'')\,,x),V] \quad (14)$$

$$L_{px} = \mathbb{E}_{px}[G(x'|p'), x] \quad (15)$$

where V is the validity matrix of ones with [n x n] dimensions. $\mathbb{E}_{GAN}$ and $\mathbb{E}_{px}$ is the error functions used, where $\mathbb{E}_{GAN}$ is the Mean Squared error function or L2 loss and $\mathbb{E}_{px}$ is calculating the absolute error or the L1 loss.
The total generator loss becomes,

$$L_G = L_{GAN} + \lambda\, L_{px} \quad (16)$$

$L_G$ is then backpropagated through the Generator model using Adam optimizer.

For the Discriminator loss, the discriminator's prediction for real and fake data are compared through the $\mathbb{E}_{GAN}$ function to generate the real and fake loss. Mathematically this can be expressed as,

$$P_{real} = D(x,\ x') \quad (17.1)$$
$$L_{real} = \mathbb{E}_{GAN}(P_{real}, V) \quad (17.2)$$
$$P_{fake} = D(G(x',p'),\ x') \quad (17.3)$$
$$L_{fake} = \mathbb{E}_{GAN}(P_{fake}, F) \quad (17.4)$$

Where V again is an [n x n] matrix of ones and F is an [n x n] matrix of zero to signify a truly fake data instance. The total Discriminator loss can be expressed as,

$$L_D = \frac{1}{2}(L_{real} + L_{fake}) \quad (18)$$

Where $L_D$ is again backpropagated through the discriminator model through Adam optimizer.

The variation in the training can be done by implementing various data input techniques during training like changing the $z_n$ distribution function by either changing the sample range or by changing the noise generator altogether. The amount of original information that is going into the GAN along with the noise can also be varied to get different kinds of outputs.

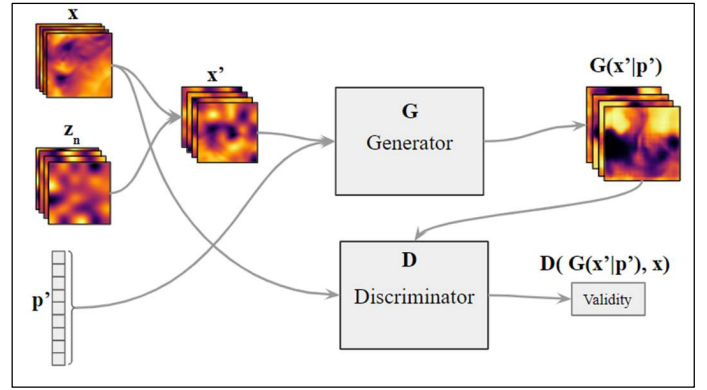

**FIGURE 4:** PIX2PIX DATA FLOW THOUGH THE GENERATOR G AND THE DISCRIMINATOR D

## 3  RESULTS AND DISCUSSION

### 3.1 Autoencoders

#### 3.1.1 Convolutional Autoencoder

Convolutional Encoder takes in the $64 \times 64$ input with 16 channels and applies convolution and max pooling layers to reduce its size and increase the channel width. After going through multiple layers, it splits into parametric and latent vector space. The decoder concatenates these to pass through linear layers similar to the encoder and finally the output with the required four channels is reconstructed through a series of UpSample layers. Other channels play their part in gradient computation. The Convolutional Autoencoder model can be more clearly understood in the figure.

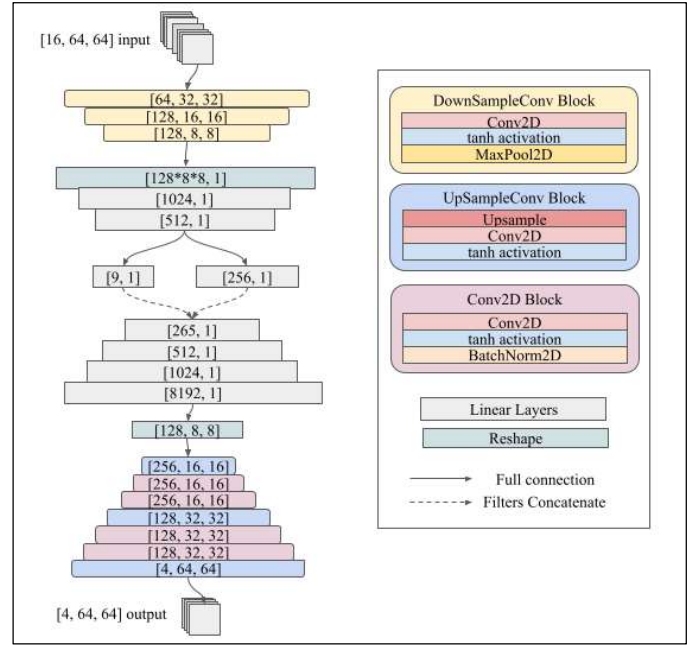

**FIGURE 5**: CONVOLUTIONAL AUTOENCODER MODEL

The summary and various hyperparameters are summarized in Table 1. Schedulers for the model and the encoder are also used to vary the learning rate after few epochs of stagnation.

Figure 10 shows the reconstruction and parameter losses for the convolutional autoencoder. Figure 8 shows the reconstruction of 5 random images from the dataset. The reconstruction seems to capture the predominant features in most cases, barring some outliers, though there is little detail in the reconstructions. Individual channel reconstructions can be seen in Figure 8.

Using the convolutional autoencoder as a generative model is demonstrated in Figure 5, which shows the linear interpolation between two images which generates new images at different parameter values.

### 3.1.2 Variational Autoencoder

Variational Autoencoder also uses convolutional and upsampling layers but works on a probabilistic principle as described earlier. Figure 7 shows the structure of the Variational autoencoder in detail. It also employs the concept of skip connections in its encoder and decoder blocks as shown in Figure 7. The summary and various hyperparameters are also summarized in Table 1.

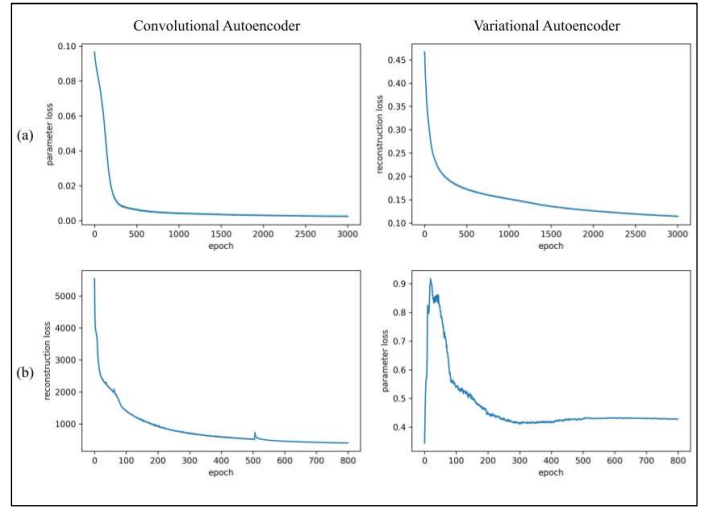

**FIGURE 6:** (a) RECONSTRUCTION AND (b) PARAMETER LOSS CURVES FOR CONVOLUTIONAL AND VARIATIONAL AUTOENCODERS

Figure 6 shows the reconstruction and parameter losses for the variational autoencoder. The reconstruction loss is the sum of losses of all individual pixels in the four channels. 5 random images from the dataset are fed into the autoencoder and the results are shown in Figure 16. The reconstructed images have good detail and capture most of the features, though some images stand out as "bad" reconstructions. Velocity and pressure channel reconstructions can be seen in Figure 8.

Interpolation in variational autoencoder results in images with moderately detailed features that smoothly transition into one another. It outperforms the convolutional autoencoder in terms of feature clarity as shown in Figure 8.

Figure 10 shows the reconstruction and parameter losses for the convolutional autoencoder. Summary of the results can be seen in Figure 8. Figure 8(a) shows the reconstruction of velocity and pressure channels of a random image from the

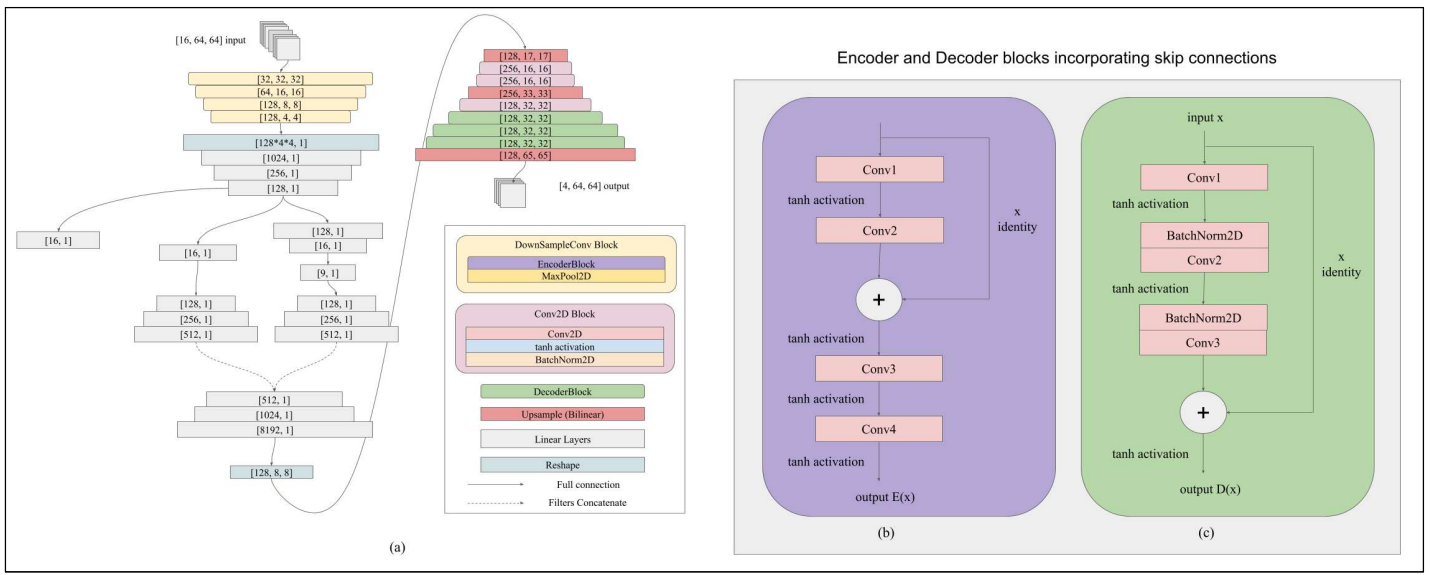

**FIGURE 7**: (a) VARIATIONAL AUTOENCODER MODEL, (b) ENCODER BLOCK AND (c) DECODER BLOCK

6    © 2021 by ASME

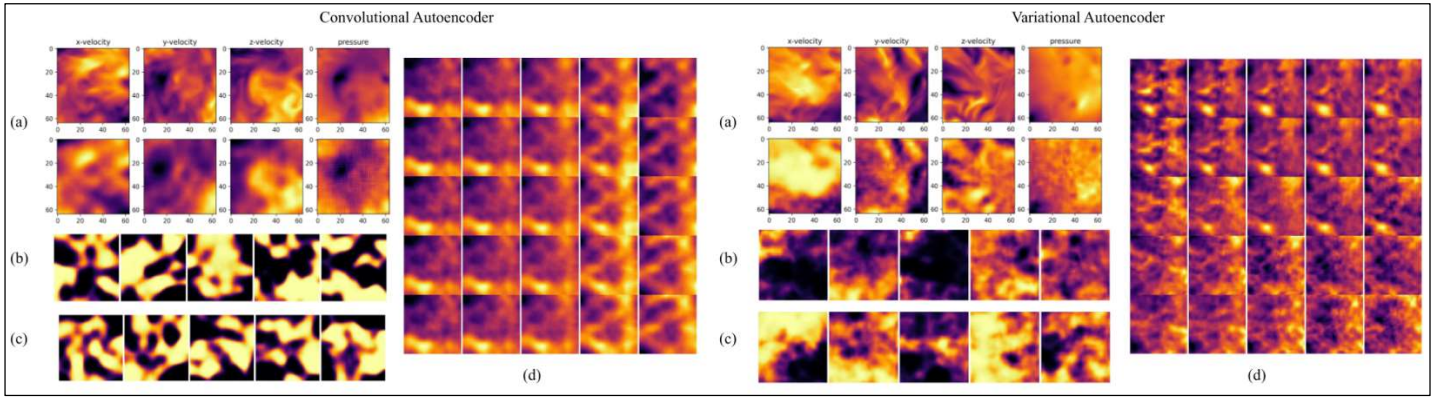

**FIGURE 8** RESULTS OF CONVOLUTION AND VARIATIONAL AUTOENCODERS. A) IMAGE RECONSTRUCTION, B) GENERATED SAMPLES (FIXED PARAMETERS), C)GENERATED RANDOM SAMPLES, D) CH:1 OUTPUT BY LINEAR INTERPOLATION OF PARAMETERS

dataset. The reconstruction seems to capture the predominant features in most cases, barring some outliers, though there is little detail in the reconstructions. Using the convolutional autoencoder as a generative model is demonstrated in Figure 8(b) and 8(c), which shows the linear interpolation between two images which generates new images at different parameter values.

**TABLE 1** TRAINING DETAILS FOR AUTOENCODER BASED MODELS

| Model | CNN-AE | VAE |
|---|---|---|
| Number of epochs | 3000 | 800 |
| Optimizer | SGD | Adam |
| Model Learning Rate | 0.001 | 0.00001 |
| Encoder Learning Rate | 0.0005 | 0.000005 |
| Batch size | 5 | 1 |
| Criterion | L1 Loss | MSE Loss |
| Activation function(s) | tanh | tanh |
| Weight decay | 1E-5 | 1E-5 |

### 3.2 Generative Adversarial models

Training generative models is a very difficult process as two losses are adversarial and are running against each other. Sometimes these models tend to become very 'explosive' and tip towards favouring either the generator or the discriminator especially in cases where the inputs and outputs have very little defined correlation and are almost like a form of noise distribution itself

### 3.2.1 Conditional GAN

The conditional GAN architecture as explained before consists of a Generator and a Discriminator. The Generator is itself a combination of MLP and CNN models. The discriminator is MLP based model which checks the validity of the input data and distinguishes it from the 'real' and 'fake' data.

The actual model architecture can be explained by the Fig 9,10.

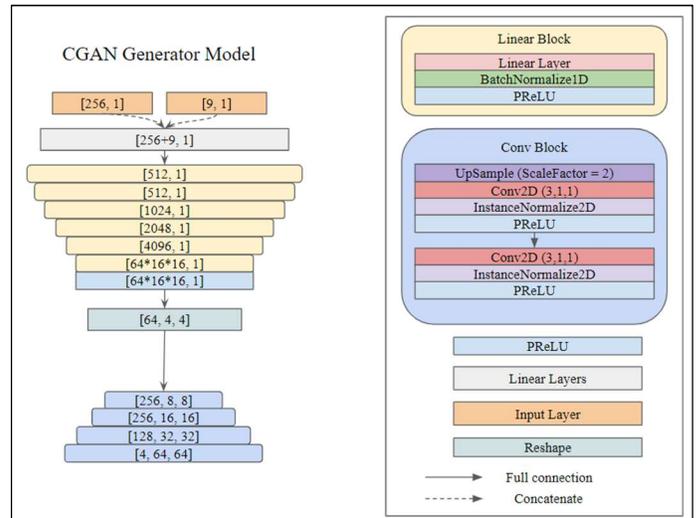

**FIGURE 9**: CGAN GENERATOR MODEL

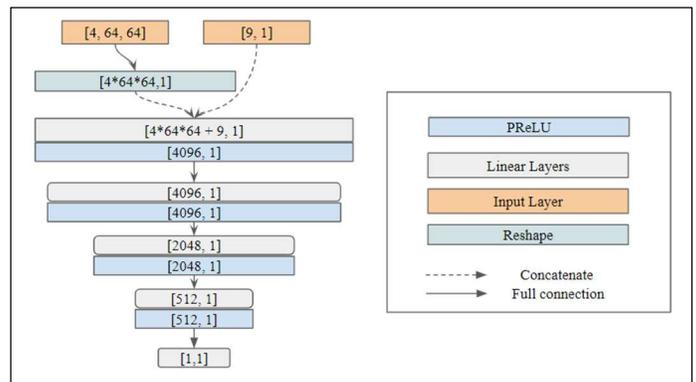

**FIGURE 10**: CGAN DISCRIMINATOR MODEL

© 2021 by ASME

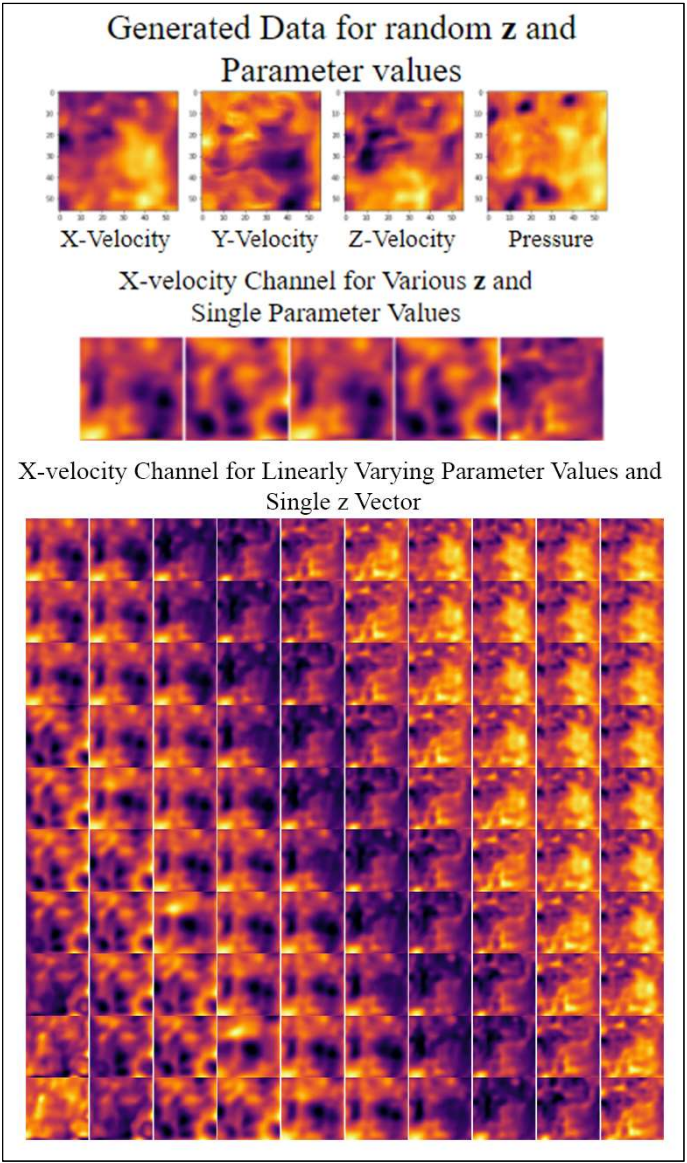

**FIGURE 11**: CGAN RESULTS

The generator outputs the x-velocity, y-velocity, z-velocity, and pressure of a single instance. The results can be seen from the Figure 11. The latent vector here is a normal distribution of 256 random numbers with a mean 0 and standard deviation of 1. The 9 flow parameters are also randomized. As it is trained on normalized parametric data, the results should mimic the pattern in the original data Fig 1.

By varying the parameter p, various conditions of flow can be created. From Figure 10, the conversion of flows into one another can be seen clearly. The overall performance of this model can be called satisfactory but not great.

### 3.2.2 pix2pix GAN Implementation

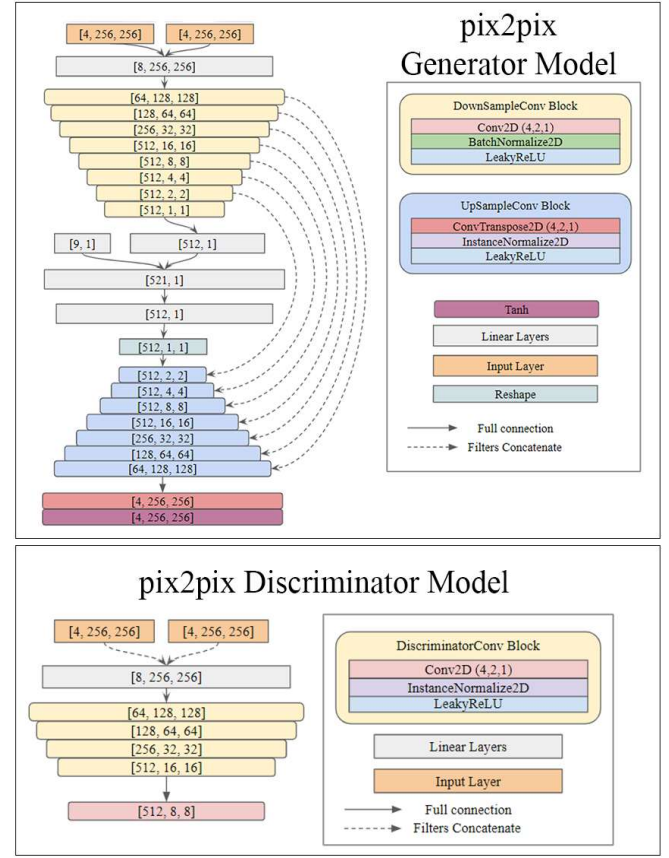

**FIGURE 12** DETAILED PIX2PIX GENERATOR AND DISCRIMINATOR MODEL WITH MENTIONED OUTPUT LAYER SIZES AND TYPES

As mentioned earlier, pix2pix architecture has been utilized to learn the flow patterns and retain information of x, y, z velocity, and pressure of various data slices and produce a new set of data by taking the original data and a 2D random distribution. The architecture of the models is given here in the figures

This kind of architecture demands 2 data inputs to transfer the 'style' from one to the other. As mentioned earlier, X and x' are the inputs of the model where x' is the noise added image. x' can be expressed as a combination of the images from the dataset x and $z_n$ which is the noise distribution, which can be expressed as

$$x' = \alpha x + \beta z_n \qquad (19)$$

Where $\alpha$ and $\beta$ are set manually to train different models. Three such training schemes have been demonstrated (Fig 13):

1. Model **M1** ($\alpha = 0$, $\beta = 1$)
   $z_n$ is a 4x8x8 grid of random values of a normal distribution with mean = 0 and standard deviation = 1. $z_n$ is upscaled to 4x256x256 by 'bicubic' interpolation.

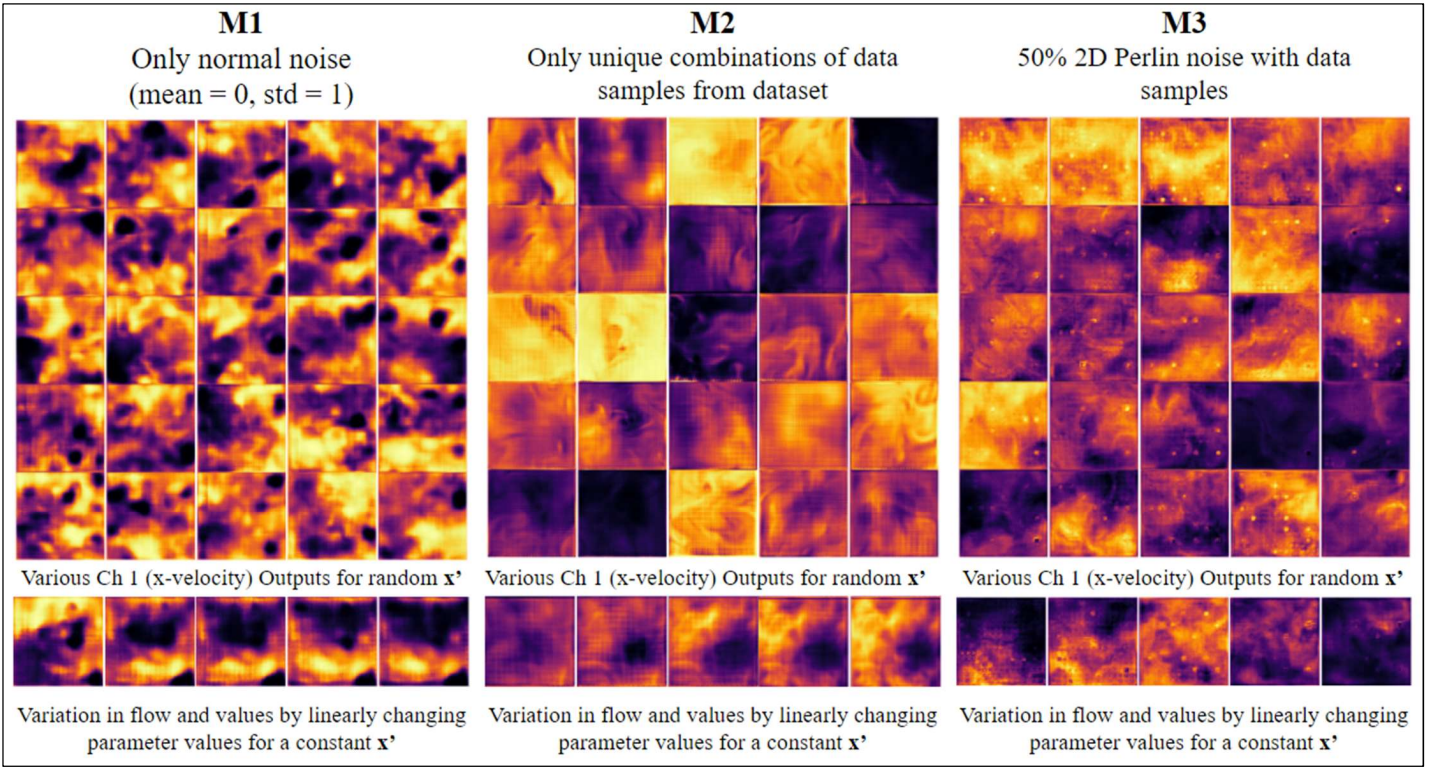

**FIGURE 13** PIX2PIX RESULTS

2. Model **M2** ($\alpha = 1$, $\beta = 0$)
   Different instances (different from the 2$^{nd}$ input) from the data are fed as the 'modified' input.
3. Model **M3** ($\alpha = 0.5$, $\beta = 1$)
   Here, $z_n$ is an 8x8 grid of 2D Perlin noise which produces procedural organic textures in the 8x8 grid for every 4 channels. $z_n$ is upscaled to 4x256x256 by 'bicubic' interpolation. 50% of additional data slices from the dataset is added to $z_n$ to form complete 'modified' input.

After adding the noise functions, it could be observed that the M1 model suffers severely from the influence of the initiation noise **x'**. M2 and M3 produce natural-looking flow velocity patterns and are comparable to the original images of the data set (Fig 1). Since M2 only had 2 different sets of inputs from the dataset itself, the generation is well defined and preserves details as well. The 'style transfer' properties of the pix2pix architecture can be seen easily. Fine flow lines and vortices can be noticed easily. M2 model suffers from low range values reproductions which signifies that the flow values are stuck in some narrow range that produces low contrast images. M3 produces excellent details and a greater value range in the generator output, but it suffers from 'artifact' creation in some cases. This can be reduced by training M3 for longer epochs.
The influence of the flow parameters can be seen on every model (Fig 13). Slight variance in the flow parameters can change the outputs in a very drastic manner.
Training details for the Generative models are shown in Table 2

**TABLE 2** GENRATIVE MODELS TRAINING SUMMARY

| Model | CGAN | pix2pix M1 | pix2pix M2 | pix2pix M3 |
|---|---|---|---|---|
| Optimizer | Adam | Adam | Adam | Adam |
| Learning rate | 0.0002 | 0.0002 | 0.0002 | 0.0002 |
| Batch size | 5 | 5 | 10 | 10 |
| Epoch | 2000 | 1500 | 1500 | 1500 |
| Generator Loss (avg) | 0.17154 | 0.46542 | 0.28369 | 0.25869 |
| Discriminator Loss (avg) | 0.48476 | 0.10318 | 0.25096 | 0.25012 |

## 4 CONCLUSION

Various autoencoder and GAN based models have been proposed and studied to reconstruct and generate natural looking fluid flows. The proposed autoencoders can be used to compress the information of the input fluid field to a few normalized parametric values which can be passed on to the decoders to generate new outputs based on the compressed parameters. In both the CNN and Variational CNN autoencoders, the decoder can work independently as a flow generator and is significantly dependent upon the 9 flow parameters extracted from the data. Similar results can be seen in both GAN implementations. CGAN generator produces comparable flow patterns to the dataset. The 'pix2pix' model which is generally used for 'style-transfer' has also proved to be very efficient in extracting information from the parameters and generating unique flows using an initiating noise distribution. The M3 implementation has proved to produce highly detailed generated flows but suffers from 'artifact creation' in some images.

 

It can be concluded that the proposed neural network based models provide a fairly less computationally expensive framework to generate and reconstruct turbulent flows based on statistical parameters. Similar models can be utilized in many areas including professional CFD applications to simulate fluid flows and in generating professional textures and effects in video game creation and VFX simulation using more elaborate 3D implementation.

Future work consists of making the strongest performing models better by more rigorous training and with a larger dataset.

## ACKNOWLEDGEMENTS

The authors would like to thank Fluid Systems Laboratory, Delhi Technological University for providing computing facilities. We are thankful to our professor Dr. Raj Kumar Singh who provided expertise which greatly assisted us in the research and improved the paper significantly. We would also like to thank other members of the Fluid Mechanics Group for directly or indirectly helping during the research work.